\def\n{\noindent}
\documentclass[twocolumn,showpacs,preprintnumbers,amsmath,amssymb]{revtex4}
\usepackage{graphicx}
\usepackage{dcolumn}

\begin{document}

\title[Vibrational properties of phonons in random binary alloys]{Vibrational properties of phonons in random binary alloys: An augmented space recursive technique in the k-representation}

\author{Aftab Alam \footnote{corresponding author}\footnote{email : alam@bose.res.in} and 
Abhijit Mookerjee \footnote{email : abhijit@bose.res.in} }
\affiliation{
     S.N. Bose National Centre for Basic Sciences, JD Block, Sector III, Salt Lake City, Kolkata 700098, India}

\begin{abstract}

We present here an augmented space recursive technique in the k-representation which include diagonal, off-diagonal and the environmental disorder explicitly : an analytic , translationally invariant , multiple scattering theory for phonons in random binary alloys. We propose the augmented space recursion (ASR) as a computationally fast and accurate technique which will incorporate configuration fluctuations over a large local environment. We apply the formalism to $Ni_{55}Pd_{45}$ , $Ni_{88}Cr_{12}$ and $Ni_{50}Pt_{50}$ alloys which is not a random choice. Numerical results on spectral functions, coherent structure factors, dispersion curves and disordered induced FWHM's are presented. Finally the results are compared with the recent itinerant coherent potential approximation (ICPA) and also with experiments.

\end{abstract} 

\maketitle

\section{Introduction}
\label{sec_Intr}

Over the years, there have been many attempts to develop adequate approximations for properties of elementary excitations in disordered systems. Among these phonons are not only conceptually the simplest, but are also the most readily accessible to experimental
verification.
 Neutron scattering experiments \cite{kambrock,ts, Nicklow} have provided detailed information about 
lattice vibrations in random alloys. A satisfactory reliable theory is still lacking. The main reason for this is that, unlike
the case of electrons in substitutionally disordered alloys, where the Hamiltonian can be expressed in a form where the disorder
is {\sl diagonal} in a real space representation (This is certainly true, for example, in a LMTO formalism in the absence of local lattice distortions), the disorder in the dynamical matrix is essentially {\sl off-diagonal}.
To make things more complicated, the diagonal  and off-diagonal disorders in  the dynamical matrix 
are coupled by the force constant sum rule  $\Phi_{RR}=-\sum_{R^{\prime}\ne R}\Phi_{RR^{\prime}}$, which ensures that 
no vibration can be excited in a uniform translation of the crystal as a whole. 
This sum rule imposes an {\sl environmental disorder} on the force constants. That is, the disorder in the diagonal
element of the dynamical matrix depends upon its near neighbours or its immediate environment. 
Hence a reliable theory for phonon excitations will be that which includes all the three kinds of disorders explicitly.

Let us look at the most successful mean field approximation : the single-site coherent potential approximation (CPA),
introduced first by Taylor \cite{Taylor}.  As the name itself suggests, it is a single site approximation and
per se cannot deal adequately with off-diagonal disorder.
Several authors have proposed schemes for generalizing the CPA  and their approaches include 
geometrically scaled off-diagonal disorder \cite{shiba,grun}, linearly scaled off-diagonal disorder \cite{kapmos} and 
independent diagonal and off-diagonal disorder \cite{kapmos,kapgray,mills,KLGD}. Most of these schemes in practice lead to a single site 
CPA including off-diagonal disorder.  These approaches suffer from two different kinds of drawbacks : first, there is no reason
why the off-diagonal part of the dynamical matrix should scale either as the geometric mean or the
arithmetic mean  of the constituents. Secondly, often the extra assumptions lead to approximate Green functions
which violate the essential Herglotz analytic properties required to produce physically acceptable results.

A complex function $f(z)$ of a complex variable $z$ is said to possess Herglotz analytic properties provided it satisfies the
following properties \cite{Mookerjee},\cite{mills} :
\begin{description}
\item[(i)] The function is analytic everywhere in the complex $z$-plane, expect on the real $z$-axis. All its
singularities therefore lie on the real $z$-axis.
\item[(ii)] The sign of the imaginary part of $f(z)$ is negative everywhere on the upper half of the $z$-plane and
positive everywhere on the lower half of the $z$-plane.
\item[(iii)] If the set of singularities of $f(z)$ is bounded, then \[ \lim_{z=E\rightarrow\pm\infty} \Re e f(E) = 0,\ \mbox{E is
real.}\]
\end{description}

The augmented space approach suggested by Mookerjee \cite{Mookerjee} provided a very interesting starting point for
the generation of appropriate approximations. The  2-site CPA proposed by 
 Yussouff and Mookerjee \cite{Yussouff}  for model systems and subsequently generalized by Mookerjee 
and Singh \cite{ms1,ms2} to 
realistic alloys was one successful approach. While it retained the Herglotz properties of the approximate Green
functions, the generalization of the 2CPA to larger clusters violated the lattice translational symmetry of the {\sl
configurationally averaged} Green function for homogeneous disorder. This was overcome subsequently by the traveling
cluster approximation of Kaplan and Gray \cite{kapgray} also based on the augmented space method.
Recently, Ghosh {\em et.al} \cite{subhra} have proposed a nearest neighbour traveling cluster CPA and have applied it 
to phonons in NiPt and NiPd alloys. In this communication we shall propose a different approximation
procedure.  We shall start 
from the augmented space method and use the recursion method of Haydock {\em et.al} \cite{hhk} to obtain the configurationally
averaged Green functions.  The termination of the continued fraction expansion will constitute the approximation.
This will not only retain the Herglotz
analytic properties of the approximate averaged Green function, but also include the effect at a site of its
neighbourhood, the size of which we can control. We shall incorporate the effect of the very distant
environment by the use of accurate {\sl termination schemes} proposed e.g. by  Haydock \cite{terminator}
, Luchini and Nex \cite{ln}
or Beer and Pettifor \cite{bp}. Since we shall incorporate the lattice translation symmetry in the
full augmented space (which is characteristic of homogeneous disorder) \cite{gdm} within our approach, the drawback of the 
original cluster-CPAs used by Singh and Mookerjee \cite{ms1,ms2} will be  overcome.
Further, we shall use the local point group symmetries of the lattice and the configurations on it to drastically
reduce the rank of the Hilbert space on which the recursion takes place (see Saha {\em et.al} \cite{sdm}). 
This will allow us to accurately account
for large environments around a particular site. One of the strengths of the proposed method which will represent a major step
forward in the theory is the possibility of including random fluctuations in force constants beyond the nearest
neighbours. While in certain representations the Hamiltonian of electronic systems can be seen to be short-ranged,
this is not so for dynamical matrices. The recursion method in augmented space can include beyond nearest neighbour
randomness in force constants without much computational expense. In our work on NiPt and NiPd we have extended 
disorder upto second nearest neighbours to illustrate this.  It is not immediately clear how easy it would be to extend
the method proposed by Ghosh {\em et.al} to larger sized clusters. We propose the ASR as a computationally fast and accurate technique which will incorporate configuration fluctuations over a large local environment. \\ 
In Section II, we shall introduce the basic formalism.  In Sections III we shall  present results for $Ni_{55}Pd_{45}$, 
$Ni_{88}Cr_{12}$ and $Ni_{50}Pt_{50}$ alloys and compare them with experiment. The choice of the systems is
deliberate : NiPd has predominantly mass disorder, NiCr predominant disorder in the dynamical matrix ; NiPt
has large disorder {\sl both} in the mass and the dynamical matrix. Concluding remarks are presented in Section IV.

\section{Formalism}
\label{sec_Form}

\subsection{\bf The augmented space formalism for  phonons}

The basic problem in the theory of phonons is to solve a secular equation of the form : 
 \[ ({\bf M}\omega^{2} - {\bf D})\ {\bf u}(R,\omega) = 0 \]  where $u_{\alpha}(R,\omega)$ is the Fourier transform of 
$ u_{\alpha}(R,t) $, the displacement of an atom from its equilibrium position $R$ on the lattice, in the direction ${\alpha} $ at time $t$. {\bf M} is the {\it mass operator}, diagonal in real space
 and {\bf D} is the {\it dynamical matrix operator} whose tight-binding representation is of the form :

\begin{eqnarray}
{\bf M} &=& \sum_{R}  m_{R}\ {\delta}_{\alpha \beta} \ P_R\\
{\bf D} &=&  \sum_{R} \Phi_{RR}^{\alpha \beta}\ P_{R} + \sum_{R}\sum_{R^{\prime} \ne R} \Phi_{RR^{\prime}}^{\alpha \beta}\ T_{RR^{\prime}}
\end{eqnarray}
along with the {\it sum rule}:
\begin{equation}  
\Phi_{RR}^{\alpha \beta} = -\sum_{R^{\prime}\ne R}\Phi_{RR^{\prime}}^{\alpha \beta}
\end{equation}

\noindent Here $P_R$ is the projection operator\ $\vert R\rangle\langle R\vert$\ \ and $T_{RR'}$ is the transfer operator\ $\vert R\rangle\langle R'\vert$\ \ in the Hilbert space ${\cal H}$ spanned by the tight-binding basis $\{\vert R\rangle\}$.
 $R,R^{\prime}$ specify the lattice sites and $\alpha $,$ \beta $ the Cartesian directions. $m_{R}$ is the mass of an atom occupying the position $R$ and $\Phi_{RR^{\prime}}^{\alpha \beta}$ is the force constant tensor. 

\noindent We shall be interested in calculating the displacement-displacement Green function in the frequency-wavevector space , which in the absence of disorder in the system has the diagonal element
\[ G({\bf {k}},{\bf {k}^{\prime}},\omega^2) = G({\bf {k}},\omega^2) \delta ({\bf {k}} - {\bf {k}^{\prime}}) \]
and for the present case
\[ G({\bf {k}},\omega^2) = \langle{\bf {k}} | \left({\bf{M}}\omega^{2}-{\bf{D}}\right)^{-1} | {\bf {k}} \rangle \]
where $| {\bf {k}} \rangle$ is a state in the reciprocal space given by :
\[ | {\bf {k}} \rangle = \frac{1}{\sqrt{N}} \sum_{R}\exp({-i{\bf{k}}.{\bf{R}}})| R \rangle \]
Since the mass matrix {\bf {M}} is perfectly diagonal, we can write 
\begin{eqnarray} 
G({\bf {k}},\omega^2)&=& \langle {\bf {k}} | {\bf{M}}^{-1/2}\left(\omega^{2}I - {\bf{M}}^{-1/2}{\bf {D}}{\bf{M}}^{-1/2}\right)^{-1}\nonumber\\
& & {\hspace {3.0cm}}\ldots{\bf{M}}^{-1/2} | {\bf {k}} \rangle
\label{eq4}
\end{eqnarray}

where \[ {\bf {M}}^{-1/2} = \sum_{R} m_{R}^{-1/2}  {\delta}_{\alpha \beta}\ P_R \]
The equation (\ref{eq4}) looks exactly like the Green function for the electronic case with ${\bf{M}}^{-1/2}{\bf {D}}{\bf{M}}^{-1/2}$ playing the role of Hamiltonian H, $\omega^{2}$ in place of energy and ${\bf {M}}^{-1/2} | {\bf {k}} \rangle$ is the starting state
of recursion.

Let us now consider a binary alloy $ A_{x}B_{y} $ consisting of two kinds of atoms A and B of masses
 $m_A$ and  $m_B$ randomly occupying each lattice sites. We wish to calculate the configuration-averaged
 Green function $\ll G({\bf {k}},\omega^2)\gg$. We shall use the augmented space formalism (ASF) to do so. Since the disorder is homogeneous, averaged $\ll G({\bf {k}},\omega^2)\gg$ is also diagonal in reciprocal space representation \cite{gdm}.  The
ASF has been described in great detail earlier \cite{tf}. We shall indicate the
main operational results and refer the reader to the above monograph for further details.
The first operation is to represent the random parts of the secular equation in terms of a random
set of local variables $\{ n_R\}$ which are 1 if the site $R$ is occupied by an A atom and 0
if it is occupied by B. The probability densities of these variables may be written as :
 
\begin{eqnarray}
 Pr(n_R)& = & x\ \delta (n_{R}-1)\ +\ y\ \delta(n_{R})\nonumber\\
 & =  & (-1/{\pi})\ \Im m\langle {\uparrow}_{R} |\ (n_{R}I-{\it {N_{R}}})^{-1}\ | {\uparrow}_{R}\rangle
\label{prob}
\end{eqnarray}

\noindent where $x$ and $y$ are the concentrations of the constituents A and B with $x+y=1$. $N_{R}$ is an operator defined on the configuration space $\phi_{R}$ of the variable $n_{R}$. This is of rank $2$ and is spanned by the states $\{|{\uparrow_{R}}\rangle, |{\downarrow_{R}}\rangle \}$ : \[ N_{R} = xp_{R}^{\uparrow} + yp_{R}^{\downarrow} + \sqrt{xy}({\tau}^{\uparrow \downarrow}_{R} + {\tau}^{\downarrow \uparrow}_{R}) \]

Let us now carry out the ASF operations in some detail.
The mass $m_{R}$ for site R can then be expressed as :
\begin{eqnarray*}
m_{R}^{-1/2} &=& m_{A}^{-1/2}n_{R} + m_{B}^{-1/2}(1-n_{R})\\
             &=& m_{B}^{-1/2} + (\delta m)^{-1/2} \  n_{R}
\end{eqnarray*}

\noindent where 

\[ (\delta m)^{-1/2} = m_{A}^{-1/2} - m_{B}^{-1/2} \]

\noindent Therefore

\begin{equation} 
{\bf M}^{-1/2} = \sum_{R}  \left[ m_{B}^{-1/2} + n_{R}\ (\delta m)^{-1/2} \right] \delta_{\alpha \beta} \ P_R
\label{M-random}
\end{equation}

\noindent In augmented space formalism, in order to obtain the configuration average we simply replace the random variables $n_R$ by the
corresponding operator $N_R$ associated with its probability density (as in equation \ref{prob}) and take the matrix element of
the resulting operator between the {\sl reference states}. The justification is sketched in the Appendix A. For a full
mathematical proof the reader is referred to \cite{Mookerjee}.\\
\begin{eqnarray*}
 n_{R}\longrightarrow N_{R} = x\ p_{R}^{\uparrow} + y\ p_{R}^{\downarrow} + \sqrt{xy}\ ({\tau}^{\uparrow \downarrow}_{R} + {\tau}^{\downarrow \uparrow}_{R}) \\
\phantom{ n_{R}\longrightarrow N_{R}} = x\ \tilde{I}\ +\ (y-x)\  p_{R}^{\downarrow} + \sqrt{xy}\ {\cal T}^{\uparrow \downarrow}_R  
\end{eqnarray*}

\noindent Using the above in equation (\ref {M-random})  we get,

\begin{eqnarray}
\widetilde{\bf M}^{-1/2} &=& m_{1}^{-1/2} \tilde{I}\otimes I + m_{2}^{-1/2}\sum_{R} p_{R}^\downarrow \otimes P_{R}\nonumber\\ 
& & \dots+\ m_{3}^{-1/2}\sum_{R}\ {\cal T}^{\uparrow\downarrow}_R \otimes P_{R}
\label{Mtilde}
\end{eqnarray}

\noindent where

\[ \left.  \begin{array}{ll}
m_{1}^{-1/2} =  x\ m_{A}^{-1/2}\ + \ y\ m_{B}^{-1/2}\\
m_{2}^{-1/2} = (y-x)\ (\delta m)^{-1/2} \\
m_{3}^{-1/2} = \sqrt{xy} \ (\delta m)^{-1/2}  \end{array} \right\} \]

\noindent Similarly the random off-diagonal force constants $\Phi_{RR^{\prime}}^{\alpha \beta}$ between the sites $R$ and $R^{\prime}$ can be written as :

\begin{eqnarray}
\Phi_{RR^{\prime}}^{\alpha \beta} &=& \Phi_{AA}^{\alpha \beta} n_{R} n_{R^{\prime}} + \Phi_{BB}^{\alpha \beta} (1-n_{R}) (1-n_{R^{\prime}})\nonumber\\
& & \ldots+ \Phi_{AB}^{\alpha \beta} \left[\ n_{R}(1-n_{R^{\prime}}) + n_{R^{\prime}}(1-n_{R})\ \right]\nonumber\\
\phantom{x} \nonumber\\
\phantom{\Phi_{RR^{\prime}}^{\alpha \beta}}  &=&  \Phi_{BB}^{\alpha \beta}\ +\ (\Phi_{AA}^{\alpha \beta} + \Phi_{BB}^{\alpha \beta} - 2 \Phi_{AB}^{\alpha \beta})\ n_{R} n_{R^{\prime}}\nonumber\\
& & \dots+ (\Phi_{AB}^{\alpha \beta} - \Phi_{BB}^{\alpha \beta})\ (n_{R} + n_{R^{\prime}}) \nonumber \\
\end{eqnarray}

\noindent Let us define the following :

\begin{eqnarray*}
\Phi^{\alpha\beta}_{(1)} &=& x\ \Phi_{AA}^{\alpha \beta} - y\ \Phi_{BB}^{\alpha \beta} + (y-x) \Phi_{AB}^{\alpha\beta}\\
\Phi^{\alpha\beta}_{(2)} &=& \Phi_{AA}^{\alpha \beta} + \Phi_{BB}^{\alpha \beta} - 2 \Phi_{AB}^{\alpha \beta}
\end{eqnarray*}

\noindent In augmented space the off-diagonal force constant matrix becomes an operator :

\begin{eqnarray}
\widetilde{\Phi}^{\alpha\beta} &=& \sum_{RR'}\ \left[\rule{0mm}{5mm} \ll \Phi^{\alpha\beta}_{RR'}\gg\ \tilde{I} + 
  \Phi^{\alpha\beta}_{(1)} \left\{ (y-x)\ (p^\downarrow_R
+p^\downarrow_{R'})\right.\right. \nonumber \\
& & \left.\left.\ldots+\ \sqrt{xy} ({\cal T}^{\uparrow\downarrow}_{R}+{\cal T}^{\uparrow\downarrow}_{R'})\right\} + \Phi^{\alpha\beta}_{(2)}\ \left\{ (y-x)^2\ p^\downarrow_R\ p^\downarrow_{R'} + \right.\right.\nonumber\\
& & \left.\left.\ldots\sqrt{xy}(y-x) \left(p^\downarrow_R\ {\cal T}^{\uparrow\downarrow}_{R'} + p^\downarrow_{R'}\ {\cal T}^{\uparrow\downarrow}_{R}
\right) + xy\ {\cal T}^{\uparrow\downarrow}_{R}{\cal T}^{\uparrow\downarrow}_{R'} \right\}\rule{0mm}{5mm} \right]\nonumber\\
& &{\hspace{6cm}}\otimes T_{RR'}
\nonumber \\
&=& \sum_{RR'}\ \Psi_{RR'}^{\alpha\beta}\otimes T_{RR'} \nonumber\\
\end{eqnarray}
\noindent The sum rule,
\[
 \Phi_{RR}^{\alpha \beta} = - \sum_{R^{\prime} \ne R} \Phi_{RR^{\prime}}^{\alpha \beta} \]
\noindent gives the diagonal element of the dynamical matrix :
\begin{equation}
\widetilde{\Phi}^{\alpha\beta} = - \sum_{R}\ \left\{ \sum_{R'\ne R}  \Psi_{RR'}^{\alpha\beta}\right\}\otimes P_R
\end{equation}

\noindent The total dynamical matrix in the augmented space is :
\begin{equation}
 \widetilde{\mathbf D} \ =\  - \sum_{R}\ \left\{ \sum_{R'\ne R} \Psi_{RR'}^{\alpha\beta}\right\}\otimes P_R \ +\ 
\sum_{RR'}\ \Psi_{RR'}^{\alpha\beta}\otimes T_{RR'}
\label{Dtilde}
\end{equation}
\\
The augmented space theorem \cite{Mookerjee} now states that the configuration averaged Green function $\ll G({\bf {k}},w^2)\gg$ may be written as :
\begin{widetext}
\begin{eqnarray}
\ll G\left({\bf {k}},\omega^{2}\right)\gg  &=&  \int G\left({\bf {k}},\omega^2,\{n_{R}\}\right)\ \prod\ Pr(n_{R})\ dn_{R}\nonumber\\
  &=&  \langle {\bf{k}} \otimes \{ \emptyset \}|\ \widetilde {G}({\bf{k}},\omega^{2},\{{N_{R}}\})| {\bf{k}}\otimes \{\emptyset\}\rangle\nonumber\\ 
& = &\langle {\bf {k}} \otimes \{ \emptyset \}|\ \widetilde{\bf {M}}^{-1/2}\left( \omega^{2}\widetilde{I} - \widetilde{\bf{M}}^{-1/2}\widetilde{\bf {D}}\widetilde{\bf{M}}^{-1/2} \right)^{-1} \widetilde{\bf{M}}^{-1/2}\ |{\bf {k}} \otimes \{\emptyset\}\rangle
\label{G_avg}
\end{eqnarray}
\end{widetext}
\noindent where $\widetilde{\bf{M}}^{-1/2}$ and $\widetilde{\bf {D}}$ are the operators 
which are constructed out of ${\bf{M}}^{-1/2}$
 and ${\bf {D}}$ by replacing all the random variables $n_{R}$ (or $n_{R^{\prime}})$ by 
the corresponding operators $N_{R}$ (or $N_{R^{\prime}})$ as given by equation  (\ref{Mtilde}) and (\ref{Dtilde}).These are the operators in the augmented space $ \Omega = {\cal H} \otimes \Phi $. The state $|{\bf {k}} \otimes \{\emptyset\}\rangle$ is actually an augmented space state which is the direct product of the Hilbert space basis and the Configuration space basis.The configuration space $ \Phi = \prod_{R}^{\otimes}\phi_{R} $ is of rank $2^{N}$ for a system of N-lattice sites with binary distribution. A basis in this space is denoted by the cardinality sequence $ \{{\cal C}\} = \{R_{1},R_{2},\ldots,R_{c}\} $  which gives us the positions where we have a $\vert\!\downarrow \rangle$ configuration. The configuration $\{\emptyset\}$ refers to a null cardinality sequence i.e. one in which we have $\vert\uparrow \rangle$ at all sites.

Using the operator representation for $ \widetilde{\bf{M}}^{-1/2} $  we  get :
\[ \widetilde{\bf{M}}^{-1/2}\ \vert {\bf {k}} \otimes 
\{ \emptyset \}\rangle \ =\ m_{1}^{-1/2} \parallel \{ \emptyset \} \rangle + m_{3}^{-1/2} \parallel \{ R \} \rangle = \vert 1 \} \]

\noindent where  a configuration state is denoted by its {\sl cardinality sequence} $\{C\}$.  We have  also used the short hand notation  :

\[ \parallel \{ C \} \rangle\ \equiv\  \frac{1}{\sqrt{N}}\sum_{R} \exp(-i \ {{\bf {k}}}\cdot R) \ | R\ \otimes \{C\} \rangle\]

\noindent  The ket $| 1 \}$  is not normalized. A normalized ket $| 1 \rangle$ is given by

\[ \vert 1\rangle = \frac{\vert 1\}}{\sqrt{\{ 1\vert 1\}} } = \left(\frac{m_1}{\widehat{m}}\right)^{-1/2} \parallel \{\emptyset\}\rangle +
\left(\frac{m_3}{\widehat{m}}\right)^{-1/2} \parallel \{R\}\rangle \]

\noindent
 With the definitions : 

\[  \ll (1/m) \gg^{-1} \ =\ \widehat{m} \]

\begin{eqnarray}
 \widetilde{\cal I} &=& \left(\frac{m_1}{\widehat{m}}\right)^{-1/2} \tilde{I} \otimes I + \left(\frac{m_2}{\widehat{m}}\right)^{-1/2}\
\sum_R  P_R \otimes p^\downarrow_R \nonumber\\
& & +\left(\frac{m_3}{\widehat{m}}\right)^{-1/2}\ \sum_R  P_R \otimes {\cal T}^{\uparrow\downarrow}_R\nonumber
\end{eqnarray}

\noindent
we may  rewrite equation (\ref{G_avg})  as :

\begin{equation}
\ll G ({\bf {k}},\omega^{2})\gg\ =\ \langle 1 \vert\ ( \omega'^{2}\widetilde{\mathbf I} - \widetilde{\bf D}_{eff})^{-1}\ \vert 1 \rangle
\end{equation} 

\noindent
 where,  ${\omega^\prime}^2\ =\ \widehat{m}\omega^2$\quad  and\quad   $\widetilde{\bf D}_{eff}$ \ =\ $\widetilde{\cal I}\ \widetilde{\bf {D}}\ \widetilde{\cal I}$\ 
 
This equation is now exactly in the form in which recursion method may be applied. At this point we note that the above expression for the averaged $\ll~{G}({\bf {k}},\omega'^{2})\gg $ is exact. The recursion transforms the basis through a three term recurrence relation as :

\begin{eqnarray}
\vert \phi_{1} \rangle = \vert 1 \rangle \quad\quad {\mathrm ;}\quad\quad  \vert \phi_0 \rangle = 0 \nonumber\\
\vert \phi_{n+1} \rangle = \widetilde{\bf D}_{eff} | \phi_{n} \rangle - a_{n} | \phi_{n} \rangle - b_{n}^2\vert  \phi_{n-1} \rangle
\end{eqnarray}

\noindent The averaged Green's function can then be written as a continued fraction :   
\[
\ll G({\bf {k}},\omega^2)\gg =
 \frac{b_1^2}{\displaystyle \omega'^{2} - a_{1} - \frac{b_{2}^{2}}{\displaystyle \omega'^{2} - a_{2}
- \frac{b_{3}^{2}}{\frac{\displaystyle \ddots}{\displaystyle \omega'^2-a_N-\Gamma({\bf k},\omega'^2)}}}}
\]
\begin{equation}
\end{equation}
\noindent where $\Gamma({\bf k},\omega'^2)$ is the asymptotic part of the continued fraction, and

\begin{eqnarray}
a_n({\bf {k}}) &=& \frac{\langle \phi_n\vert \widetilde{\mathbf D}_{eff}\vert \phi_n\rangle}{\langle \phi_n\vert \phi_n\rangle}\quad \mbox{ and }\quad 
\nonumber\\
b_n^2({\bf {k}}) &=& \frac{\langle \phi_{n}\vert  \phi_n\rangle}{\langle\phi_{n-1}\vert\phi_{n-1}\rangle} \quad \mbox{ ; }\quad 
b_1^2 = 1 \nonumber
\end{eqnarray}
\begin{equation}
\end{equation}

To implement the above recursion, we require to know the effect of the operator $\widetilde{\bf D}_{eff}$ on a general state in augmented reciprocal space \cite{k-space}. Some of the main operations are shown in Appendix B.

So far the expression for the averaged Green function is {\sl exact}. Approximations are introduced at this stage for
its actual numerical evaluation. The mean-field theories essentially obtain the self-energies because of disorder
scattering, self-consistently and  approximately and then calculate the averaged Green function either from the Green function without disorder or the virtual crystal Green function. The Coherent potential approximation (CPA) proposed by Soven  \cite{sov}, 
the cluster CPA proposed by Mookerjee and Singh \cite{ms1,ms2}, the traveling cluster approximation (TCA)
proposed by Mills and Ratanavaraksa \cite{mills} and Kaplan {\em et.al} \cite{KLGD} and the itinerant CPA (ICPA) proposed by Ghosh {\em et.al} \cite{subhra} basically all belong to this
category. The latest work referenced represent the most sophisticated version of the mean-field theories.
We shall propose an approximation that will start from the infinite continued fraction and approximate
its asymptotic part by an  
  analytic termination procedure. The coefficients $a_{n}$, $b_{n}$ 
are calculated exactly up to a finite number of steps and the asymptotic part is then replaced by a terminator :
$\Gamma({\bf k},\omega'^2)\simeq T({\bf k},\omega'^2)$. 
The concept of terminators is described in the appendix C, where further details of estimating  $T({\bf k},\omega'^2)$
 are also provided.
Haydock and co-workers \cite{terminator} have carried out extensive studies of the errors 
involved and precise estimates are available in the literature. 
Several terminators are available and we have chosen to use that of  Luchini and Nex \cite{ln}. 
If we calculate the coefficients up to the $n$-th step exactly the first $2n$ moments of the density of
states are reproduced exactly. The terminator ensures that the approximate Green function has Herglotz analytic
properties which also tells us that the approximate density of states is always positive definite and the 
spectrum is always real. The terminator is also so chosen that the asymptotic moments are also accurately
reproduced. This is a generalization of the method of moments, with the additional restriction that the
asymptotically large moments are also accurately obtained.

\noindent In the absence of  disorder in the problem, the Green function for a given mode is of the form :

\[ G_0({\bf {k}},\omega'^2) = \frac{1}{\omega'^2 - \omega_0^2({\bf {k}})} \]

The spectral function $A_0({\bf {k}},\omega'^2)$ is a delta function of the form $\delta(\omega'^2 - \omega_0^2({\bf {k}}))$. If we write :
\begin{eqnarray}
\Sigma({\bf {k}},\omega'^2)& = & 
 a_1({\bf {k}}) - \omega_0^2({\bf {k}}) +  \frac{b_{2}^{2}({\bf {k}})}{\displaystyle \omega'^{2} - a_{2}({\bf {k}})
- \frac{b_{3}^{2}({\bf {k}})}{\displaystyle \phantom{xxx}\ddots}} \nonumber\\
& = & a_1({\bf {k}}) - \omega_0^2({\bf {k}}) + \sigma({\bf {k}},\omega'^2)
\end{eqnarray}

\noindent Then,

\[ \ll G({\bf {k}},\omega'^2)\gg \ =\ G_0\left({\bf {k}}, \omega'^2-\Sigma({\bf {k}},\omega'^2)\right)\]

\noindent Obviously from above $\Sigma({\bf {k}},\omega'^2)$ is the disorder induced self-energy.
Damped vibrations occur with reduced frequencies at  $\omega'_0({\bf {k}})$ which are the
solutions of the implicit equation~:

\[ {\omega'_0}^2({\bf {k}})\ -\ a_1({\bf {k}})\ -\ \Re e\ \sigma({\bf {k}},{\omega'_0}^2({\bf {k}})) = 0\]

\noindent and their disorder induced widths are :

\begin{equation}
{\cal{W}}({\bf {k}},{\omega'_0}^2({\bf {k}})) = -\ \frac{1}{\pi}\ \Im m\ \sigma({\bf {k}},{\omega'_0}^2({\bf {k}}))\end{equation}

The average  spectral function  $  \ll A_\lambda ({\bf k}, w^{2})\gg
 $ for a mode labelled $\lambda$ is related to the averaged Green function in reciprocal space as : 

\begin{equation}
\ll A_\lambda({\bf k}, {\omega'}^{2})\gg  = -\frac{1}{\pi} \lim_{\delta \to 0^{+}} \left[\rule{0mm}{4mm} \Im m \left\{  \ll {G}_\lambda({\bf k}, 
{\omega'}^{2} - i \delta)\gg \right\} \right]
\end{equation}

The averaged density of states is given by :

\[ \ll n(\omega')\gg = \frac{2\omega'}{3}\ \sum_\lambda\ \int_{BZ} \ \frac{d^3{\bf {k}}}{8\pi^3}\ \ll A_\lambda ({\bf {k}},{\omega'}^2) \gg \]

\noindent Here $\lambda$ labels the particular normal mode branch and  BZ is the Brillouin zone.

The dispersion curves for different modes are then obtained by numerically calculating the peak frequencies of the spectral function.
This averaged spectral function gives, in principle, a proper description of the dynamics but it 
does not involve any weighting by scattering lengths. 
The dispersion curves so obtained are nearly the same as those obtained  experimentally from the peak frequencies 
of the coherent structure factors $S_{coh}$. This is because the coherent structure factors are nothing but the averaged
Green functions   weighted by the coherent scattering lengths.  The intensities and the line shapes measured from  $\ S_{coh}\ $  and the imaginary part of Green function may differ significantly, but the peak positions will generally differ little.

\subsection{The coherent scattering structure-factors}

Experimental determination of the phonon dispersion and line-widths are deduced from the averaged coherent scattering 
structure factors.
The expression for these can be written as~:

\begin{equation}
\ll S_{coh}({\bf k},\omega^2)\gg   = 
 -\frac{1}{\pi} 
 \Im m \ll {\mathbf b}\ {\mathbf G}({\mathbf q},\omega^2)\ {\mathbf b}\gg
\end{equation}

\n here, thermal neutrons with wave-vector {\bf k} gets scattered to final state of wave-vector {\bf k$^\prime$}, {\bf q} =
{\bf k}--{\bf k$^\prime$}+{\bf Q} with {\bf Q} being a reciprocal lattice vector. The energy lost by the incoming neutrons
are taken up by the phonons : $(\hbar^2/2M_n) (k^2-k'^2)\ =\ \hbar\ \omega$. $f(\omega)$ is the Bose distribution function and  

\[{\bf b}= \sum_R b_R\ \delta_{\alpha\beta}\ P_R\] where $b_R$ is the scattering length of the nucleus occupying the
site $R$. This  is a random variable taking two values $b_A$ or $b_B$ depending on which kind of
atom sits at the site labelled $R$. For comparison with experiment we have to calculate  $ -(1/\pi)\ \Im m\ 
\ll \left(\rule{0mm}{3mm}{\bf b\ G}({\mathbf q},\omega)\ {\bf  b}\right)^{\alpha\alpha}\gg$, rather than the spectral function. For ordered materials the two
are proportional, but if the scattering lengths are themselves random then although this has very little effect on the
dispersion curves, it does affect the line-shapes and line widths. We can easily implement such an average within the ASF~:

\begin{widetext}
\[
\ll {\mathbf b}\ {\mathbf G}({\mathbf q},\omega^2)\ {\mathbf b} \gg\ =\   \langle {\mathbf q}\otimes\{\emptyset\}\vert\ \tilde{\mathbf{b}}
\ \left( \widetilde{\mathbf M}\omega^2\ -\ \widetilde{\mathbf D}\right)^{-1}\ \tilde{\mathbf b}\ \vert {\mathbf q}\otimes\{\emptyset\}\rangle \]
\end{widetext}

\n where 

\begin{eqnarray}
\tilde{\mathbf b} &= & \ll {\mathbf b}\gg \tilde {I}\otimes I + (y-x)(b_B - b_A) \sum_{R} p^\downarrow_R\otimes P_R\nonumber
\\
& & \ldots + \sqrt{xy}(b_B-b_A) \sum_R\ {\cal T}^{\uparrow\downarrow}_R\otimes P_R\nonumber 
\end{eqnarray}

\n Carrying out algebra similar to the one for the averaged Green function, we obtain :

\begin{equation}
\ll {\mathbf b\ G}\left({\mathbf q},\omega^2\right)\ {\mathbf b}\gg \ =\ \langle 1_b\vert\ \left(\omega_{b}'^2\tilde{{\mathbf I}}-\widetilde{{\mathbf D}}^b_{eff}\right)^{-1}\ \vert 1_b\rangle
\end{equation}

\n where,

\[ \vert 1_b\rangle = \left(\frac{X_1}{\hat{X}}\right)^{-1/2} \parallel \{\emptyset\}\rangle + 
\left(\frac{X_3}{\hat{X}}\right)^{-1/2} \parallel \{R\}\rangle \]

\n with,

\begin{eqnarray}
X_1^{-1/2} & = & x\ m_A^{-1/2}b_A + y\ m_B^{-1/2}b_B \nonumber\\
X_2^{-1/2} & = & (y\ -\ x)\left( m_A^{-1/2}b_A - \ m_B^{-1/2}b_B\right)\nonumber\\
X_3^{-1/2} & = & \sqrt{xy}\left( m_A^{-1/2}b_A - \ m_B^{-1/2}b_B\right)\nonumber\\
\hat{X} & = & \ll\frac{b^2}{m}\gg \ =\ x\ \frac{b_A^2}{m_A}\ + y\  \frac{b_B^2}{m_B}\nonumber\\
\end{eqnarray}
\n Also

$\omega_{b}'^2\ =\ \hat{X}\omega^2$\ \ and \ \ $ \widetilde{\mathbf D}^b_{eff}\ =\  \widetilde{W}\ \widetilde{\mathbf D}\ \widetilde{W} $

\n where,

\begin{eqnarray}
 \widetilde{W} &=& \left(\frac{m_1}{\widehat{X}}\right)^{-1/2} \tilde{I} \otimes I + \left(\frac{m_2}{\widehat{X}}\right)^{-1/2}\
\sum_R  P_R \otimes p^\downarrow_R \nonumber\\
& & \ldots+\left(\frac{m_3}{\widehat{X}}\right)^{-1/2}\ \sum_R  P_R \otimes {\cal T}^{\uparrow\downarrow}_R
\end{eqnarray}

The subsequent recursion calculation follows the identical steps as for the averaged spectral functions. We have chosen a second neighbour force constant model, with dynamical matrices fitted to reproduce the dispersion curves.  The disorder induced widths are the quantities which are more sensitive to the effect of randomness as compared to the frequencies (i.e.\ dispersion curves\ ), and as such will be one of the focus of this work. In order to extract the full width at half maxima (\ FWHM\ ), we have fitted the coherent structure factors to Lorenzians exactly as the experimentalists do to extract the same. The advantage of including the scattering length fluctuation will be clear when we will show the nature of the line widths for $Ni_{55}Pd_{45}$ alloy with and without inclusion of the scattering length fluctuation (i.e.\ Calculating the widths once by fitting the spectral functions to Lorenzians and then the structure factors to Lorenzians\ ).  Our aim in this communication is to propose the augmented space recursion as a useful technique  to study effects of diagonal, off-diagonal and environmental disorder. Accurate model building or obtaining the force constants from first-principles total energy calculations will be postponed for future work.

\begin{table}
\centering
\vskip 0.5cm
\begin{tabular}{|l|c|c|c|r|}\hline
          &   Ni   &   Pd   &   Pt   &   Cr  \\ \hline
Atomic number    &28&46&78&24\\ \hline
Atomic mass (amu)   &58.71&106.4&195.09&51.996\\ \hline
Free atom valence config.   &$3d^{8}4s^{2}$&$4d^{10}$&$5d^{9}6s^{1}$&$3d^
{5}4s^{1}$\\ \hline
 Lattice constant (fcc)    &3.524&3.8904&3.924&3.68(fcc) \\
                            &&&&                2.89(bcc) \\ \hline
Elastic constants at &&&& \\
$296^{o}K$  ($10^{12} dyne/cm$) &&&&  \\ \hline
\ \ \ \ \ \ \ \ \ \ \ \ \ $C_{11}$    &2.461&2.270&3.467&3.5\\ \hline
\ \ \ \ \ \ \ \ \ \ \ \ \ $C_{12}$    &1.501&1.759&2.507&0.678\\ \hline
\ \ \ \ \ \ \ \ \ \ \ \ \ $C_{44}$    &1.220&0.717&0.765&1.010\\ \hline
n-n force constants &&&& \\
   (in units of dyne/cm) &&&& \\ \hline
\ \ \ \ \ \ \ \ \ \ \ \ \  1XX    &17319&19337&26358&37483\\ \hline
\ \ \ \ \ \ \ \ \ \ \ \ \  1XY    &19100&22423&30317&17453\\ \hline
\ \ \ \ \ \ \ \ \ \ \ \ \  1ZZ    &-436&-2832&-7040&-13229\\ \hline
n-n-n force constants &&&& \\
   (in units of dyne/cm)  &&&& \\ \hline
\ \ \ \ \ \ \ \ \ \ \ \ \  2XX    &1044&1424&4926& - \\ \hline
\ \ \ \ \ \ \ \ \ \ \ \ \  2YY    &-780&210&-537& - \\ \hline
\colrule
\end{tabular}
\caption{ General properties of fcc  Ni, Pd, Pt and Cr. The
force constants for Ni, Pd and Pt are taken from \cite{forceconst}
 and that for Cr is taken from \cite{ms2} }
\label{table1}
\end{table}

\begin{figure*}
\includegraphics[width=8cm,height=3.2cm]{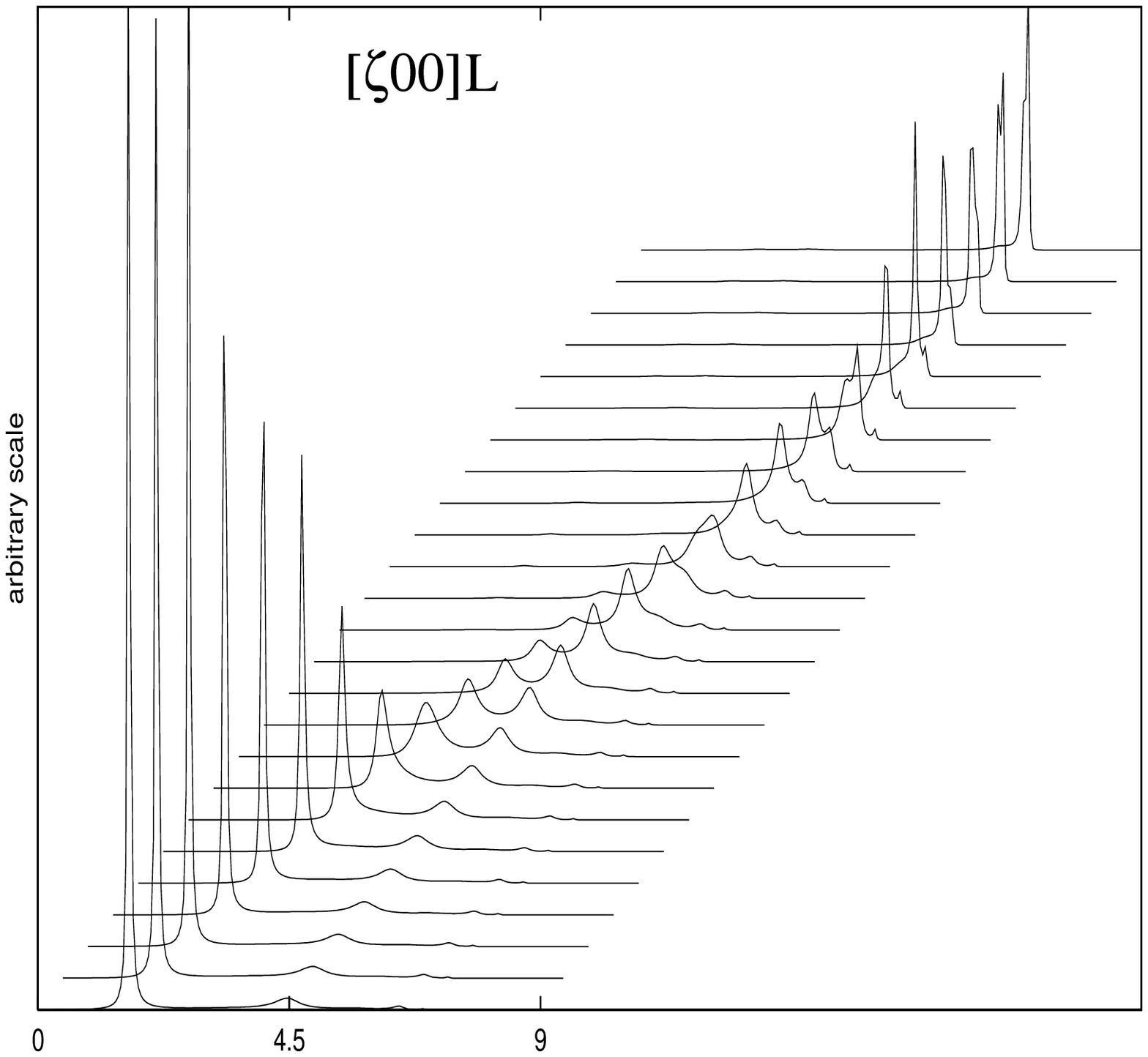}
\includegraphics[width=8cm,height=3.2cm]{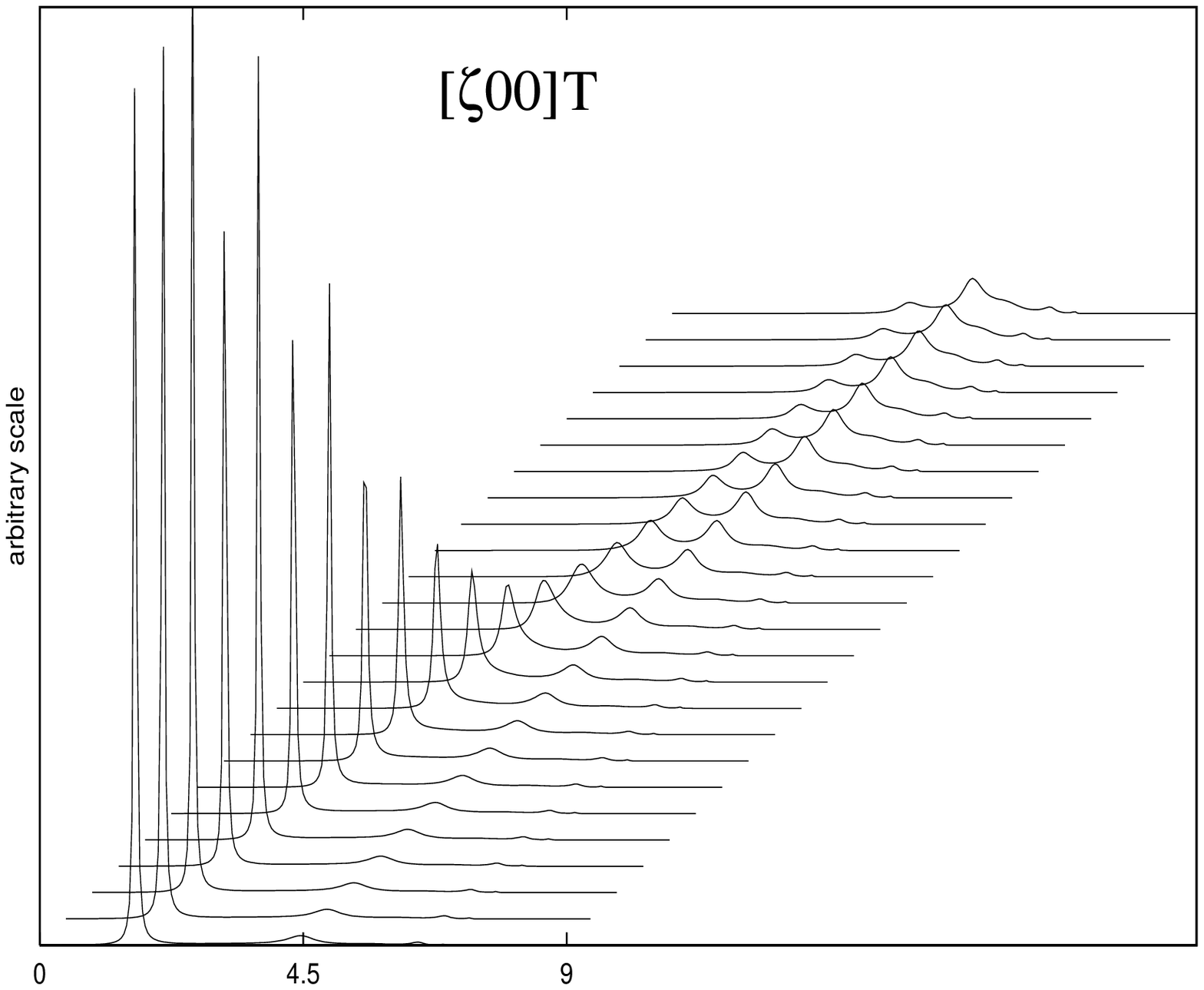}
\includegraphics[width=8cm,height=3.2cm]{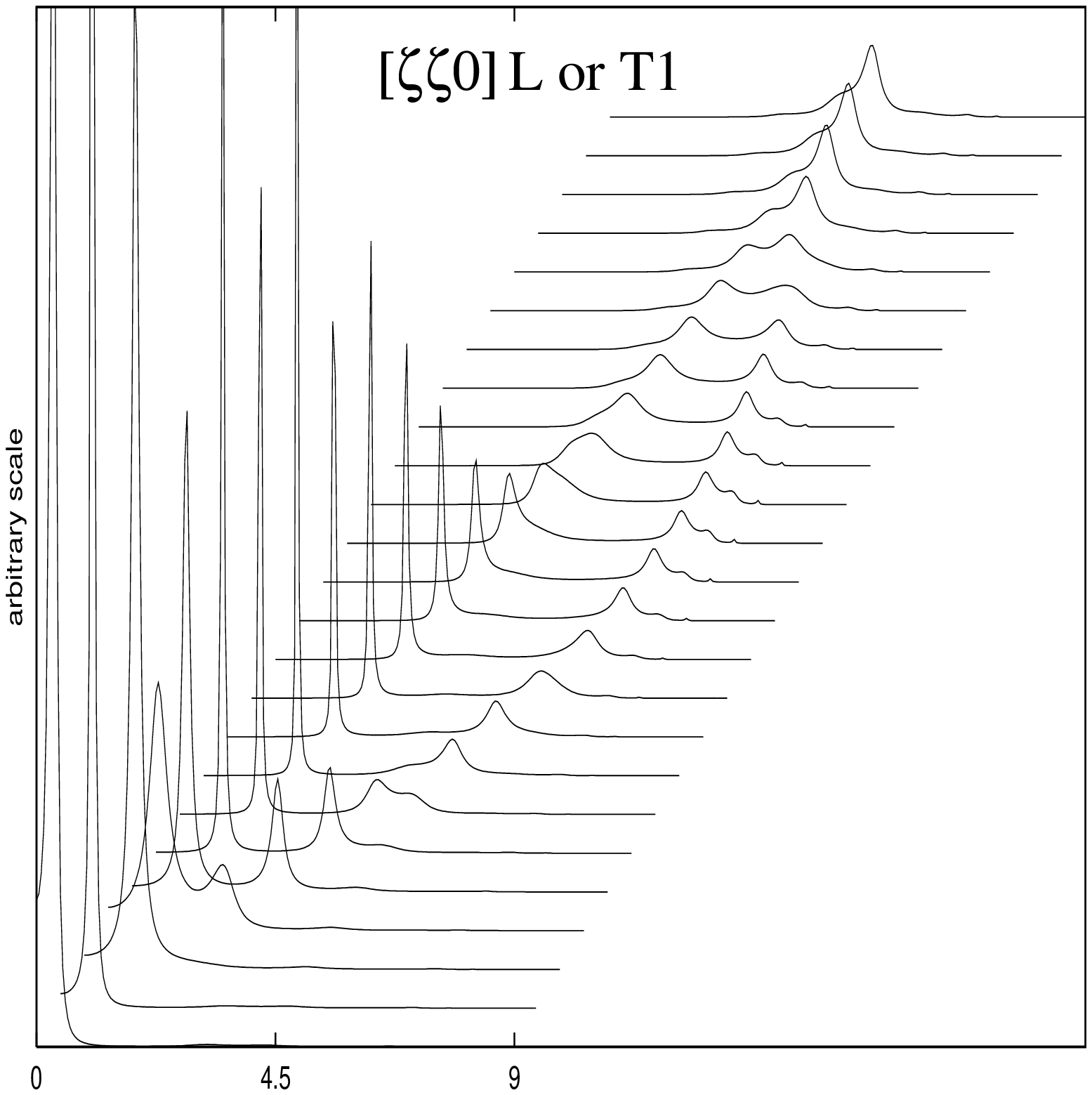}
\includegraphics[width=8cm,height=3.2cm]{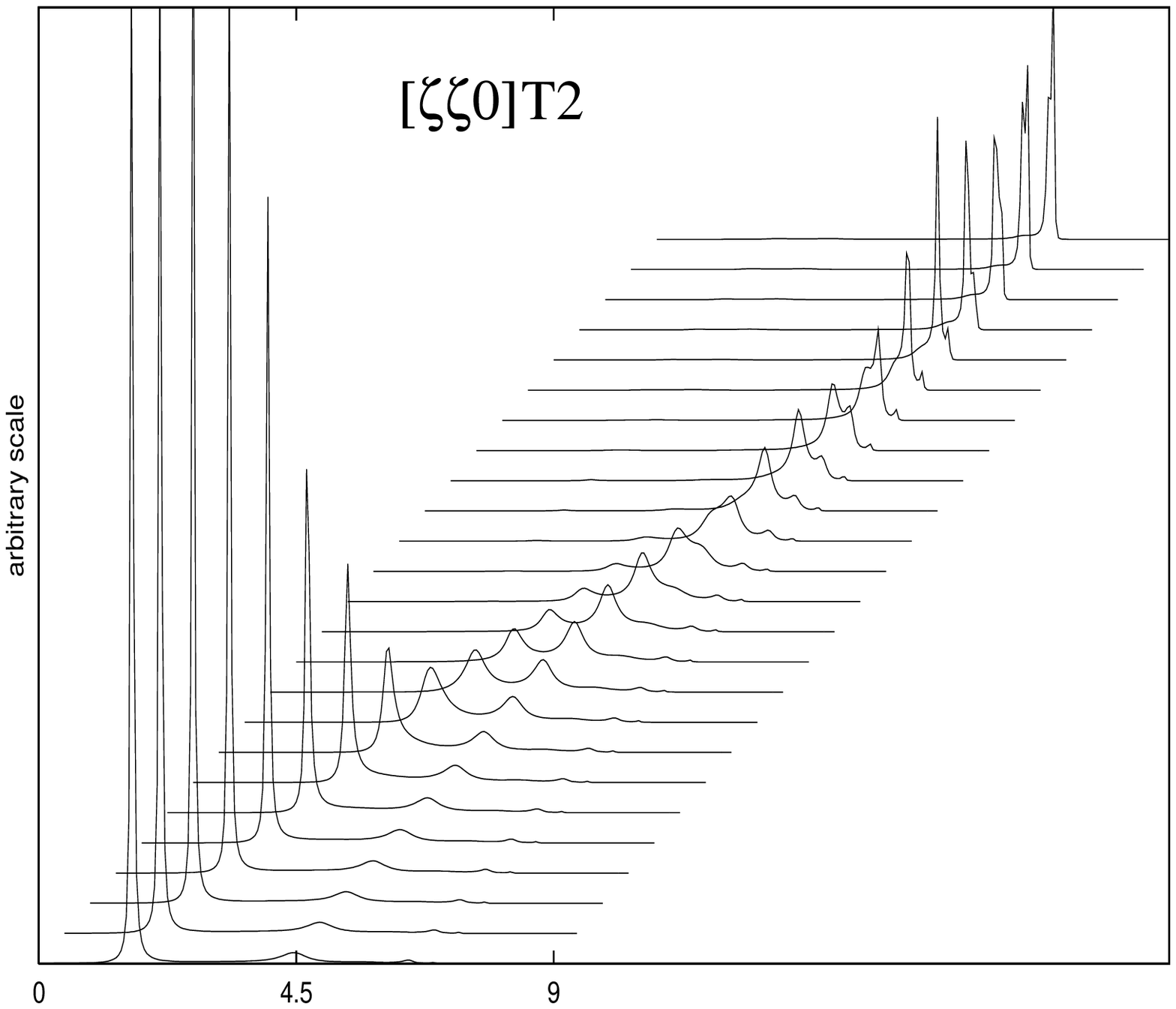}
\includegraphics[width=8cm,height=3.2cm]{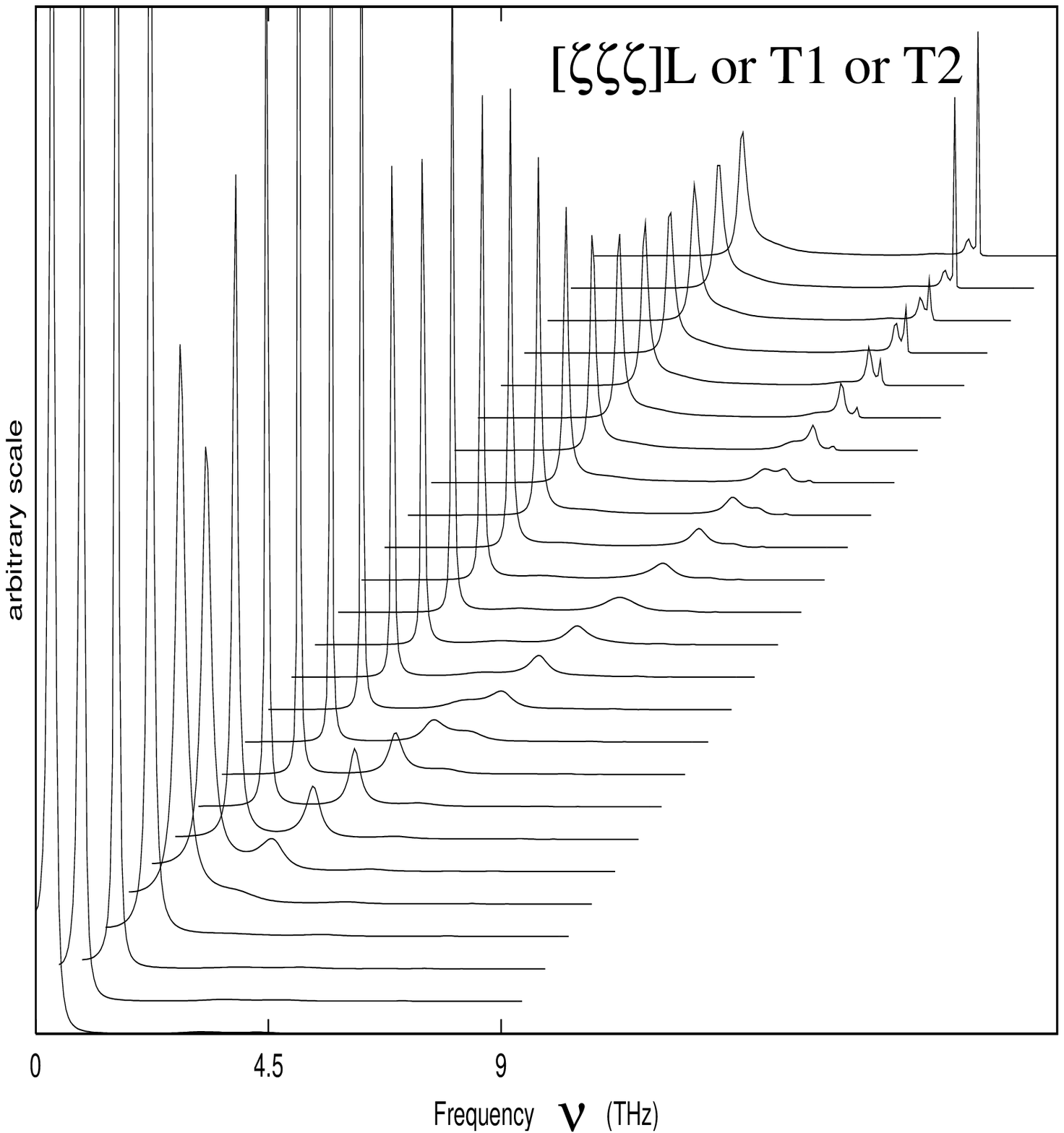}
\caption{Total coherent structure factors in different directions with different branches for Ni$_{55}$Pd$_{45}$ . In each of the different directions and branches, the various curves indicate the total structure factors for various $\zeta$ values starting from the lowest value to the edge of the Brillouin zone. In [$\zeta$ 0 0] direction T1 and T2 modes are degenerate, in [$\zeta$ $\zeta$ 0] direction L and T1 modes are degenerate and in [$\zeta$ $\zeta$ $\zeta$] direction all the three modes are degenerate. The y-axis is in an arbitrary
scale with heights scaled to the maximum height. Different curves for different $\zeta$ values are shifted along the x-axis in order to facilitate vision.}
\label{specNiPd}
\end{figure*}

In the following three Subsections, we present our calculations on
$Ni_{55}Pd_{45}$, $Ni_{88}Cr_{12}$ and $Ni_{50}Pt_{50}$ alloys. The choice is not arbitrary. Mass disorder
dominates in NiPd alloys while force constant disorder is large in NiCr alloys.  NiPt alloys have large disorder {\sl both} 
in mass and force constants. Since in the phonon problem we have both kinds of disorder, it would be interesting
to note the interplay between them in this series of alloys.  The concentrations are chosen so that we may compare our results with existing work.

\section{Results and Discussion}

\subsection{The {Ni$_{55}$Pd$_{45}$\ Alloy} }

A look at Table~\ref{table1} immediately shows us that for NiPd alloys, the dominant disorder is in the mass. Force constants in Pd are only about 15\% larger than those in Ni.
We shall choose the Ni$_{55}$Pd$_{45}$ alloy for the application of our formalism developed in section II. This
particular alloy has already been studied within the ICPA by Ghosh {\em et.al} \cite{subhra}, CCPA by Mookerjee
and Singh \cite{ms2}   and experimentally \cite{kambrock} by inelastic neutron scattering. 
 The alloy forms a continuous series of face centered cubic  solid solutions of all concentrations 
and there are no indications of long range order down to $0^{o}C$. 

In Figure {\ref {specNiPd}}  we display the coherent scattering structure factors obtained from our
 recursion calculation along the highest symmetry directions ( [$\zeta 0 0$], [$\zeta\zeta 0$],
 [$\zeta\zeta\zeta$] ), \ $\zeta = {|\vec{k}|}/{|\vec{k}_{max}|}$ for different branches. For a particular direction and branch the different curves indicates the spectral functions for various $\zeta $ points starting from the lowest value (~i.e.\ $\zeta = 0 $\ ) to the edge of the Brillouin zone (\ i.e.\ $\zeta = 1 $\ in units of\ $2\pi/a$\ ). 
The first thing to note is that the structure factors are (in contrast to Lorentzian shape) often asymmetric near the resonances. The asymmetries can be described as a tendency of more scattering to occur near the resonance frequencies. 
In other words the shape of a mode with a frequency slightly lower or higher than that of a resonance  tends to have a second peak or wide tail over the resonance region. 
In fact if one looks at the [$\zeta\zeta 0$] L or T1 ( doubly degenerate ) and [$\zeta\zeta\zeta$] L or T1 or T2
  (3-fold degenerate) branches, the shape of a doubly peaked structure factor is much more clear. 
Out of these two peaks, one peak corresponds to the dispersion curve for the longitudinal mode (L) 
and the other peak to the transverse mode (T). 
Experimentally, for some neutron groups corresponding to transverse phonons with frequencies just below the lower resonance, definite asymmetry to the right was observed. Such asymmetries are clearly observed for the
[$\zeta$ 00]T  and [$\zeta\zeta 0$]T1 branches.
It is important to note that the structure factors have a pronounced ${\bf k}$ and branch dependence.

\begin{figure}
\includegraphics[width=8.5cm,height=6cm]{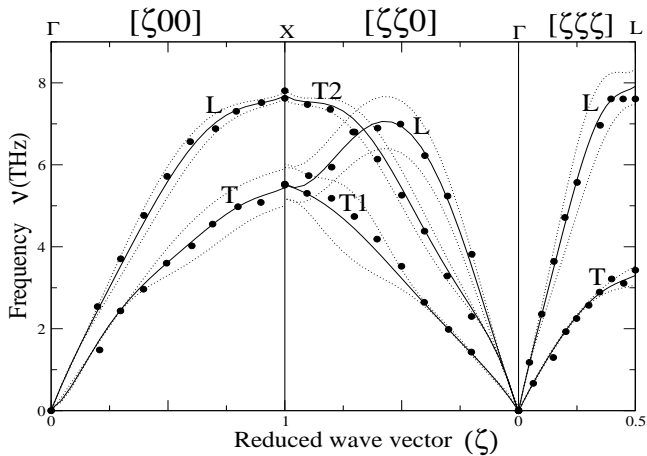}
\caption
{Dispersion Curves (\ frequency $\nu$ vs. reduced wave vector $\zeta$\ ) for Ni$_{55}$Pd$_{45}$ calculated from recursion (solid line). The force constants used are given in the text. The filled circles are the experimental data \cite{kambrock} . In all the three panels the thin dotted lines span the FWHM's . }
\label{dispNiPd}
\end{figure}

In figure \ref {dispNiPd} we display the dispersion curves, which were constructed by numerically determining the peaks in the coherent scattering structure factor.
In this communication our main focus is the development of the augmented space recursion method. Accurate
determination of the force constants shall be left for the future. Ghosh {\em et.al} \cite{subhra} have attempted
 much more detailed determination of the force constants. For the time being we have used the same parametrization of the force constants as they did. These dispersion curves (\ solid lines\ ) are compared with the experimental results \cite{kambrock} (\ filled circles\ ). The dotted lines span the calculated FWHM's. 
The procedure of calculating FWHM's has already been discussed.  The asymmetry in the widths is again
clearly observed in the two transverse branches quoted above.
The results are in good agreement with the experiment for all the three symmetry directions
 and for each branch.  Agreement can be achieved by varying only one parameter in the force constant
matrix. 
This suggests that the force constant disorder is weak and the system is dominated by the mass disorder,
 as is clear from the numerical values of the parameters given in Table~\ref{table1} . 
If one looks at the previous results for the dispersion curves (\ i.e.\  VCA, CPA and ICPA curves\ ) \cite{subhra}, it will be clear that in the low wave vector regime, there is no distinction between these results 
and ours, Because the self averaging of both mass and force constants over a single wavelength reduces the CPA, ICPA and the ASR results to become close to the VCA. 
However as we move toward high wave vectors, the VCA curve deviates from the experimentally observed one and lies
 lower in frequency as compared to the ICPA and the ASR results. The reason is that VCA uses an averaged mass. In contrast to this, for those theories which capture the effect of mass fluctuation (as do the ICPA and
the ASR), the lighter atoms (Ni in this case) dominate in the high wave vector region and push the frequencies up. That is why our results agree very well across the Brillouin zone.

\begin{figure*}
\includegraphics[width=15cm,height=7.5cm]{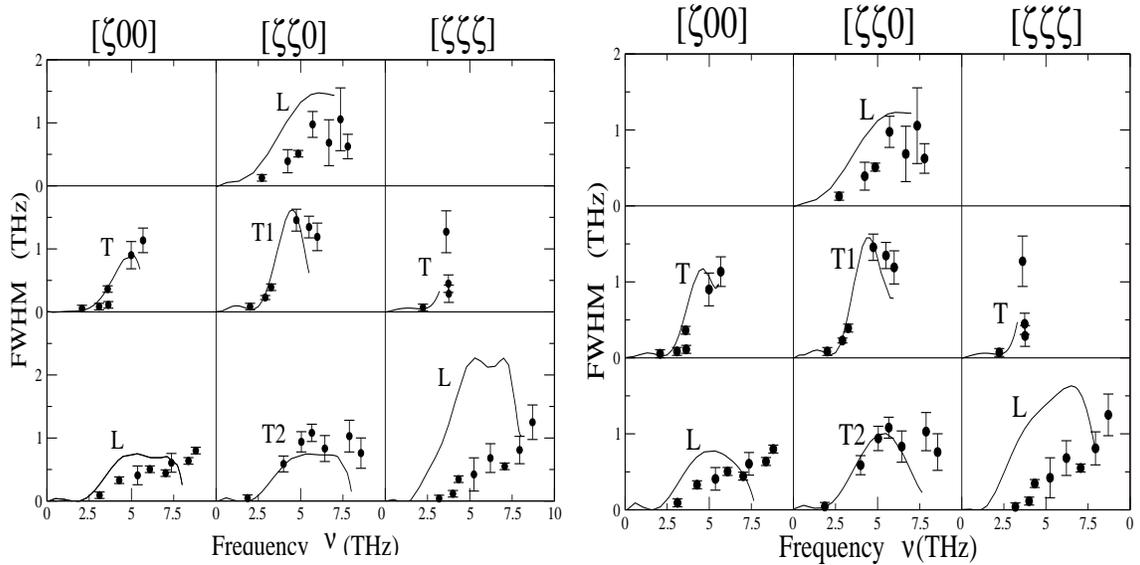}
\caption
{Full widths at half-maximum for the NiPd alloy as function of frequency for different directions
in k-space and different modes without (left)  and with (right) the inclusion of the scattering length fluctuation.The filled circles along with the error bars are the experimental data \cite{kambrock}}.
\label{widNiPd}
\end{figure*}

The FWHM's are  much more sensitive to approximations as compared to the frequencies. These are shown in figure (\ref{widNiPd}). The FWHM's shown in the left are those which have been calculated without including any scattering length fluctuation, while in the right are those where the fluctuation has been included. The circles along with the error bars are the experimental data \cite{kambrock}. It is obvious that the nature of the line widths are not the same in the two cases, rather the one including scattering length fluctuation is matching more closely with the experimental data than the one without including the fluctuation. That should be obvious because the experimentalists do include this fluctuation. Our results show very strong branch and wave-vector dependent
widths and in good agreement with the experimental results of Kamitakahara and Brockhouse except in the [$\zeta\zeta\zeta$]L mode. The reason for this may be because of the highly asymmetric line shapes in the [$\zeta\zeta\zeta$]L mode. The single site CPA
yields branch and {\bf k} independent widths. It cannot capture the essentially off-diagonal disorder
of the problem. The ICPA and the ASR manages to capture this feature. One should note that 
 the structure factors are often asymmetric in shape and the usual Lorenzian fits carried out by
most people may not be valid.  
 
\subsection{The Ni$_{88}$Cr$_{12}$ alloy}

We shall choose this alloy as being the nearest to that studied experimentally by Bosi {\em et.al} \cite{Bosi}.
Determination of the force constant matrices for this alloy becomes difficult, because pure Cr is body
centered cubic, but alloyed with Ni, up to 30$\%$ Cr it forms face centered cubic alloys. The force constants of
pure Cr may be nothing like those of Cr in this alloy. Until we are able to determine these from a more
first-principles type approach, our determination of the force-constants for this alloy will remain
tentative. We shall consider a hypothetical fcc Cr, whose force constants are related to the elastic constants of bcc Cr via :

\begin{eqnarray*}
C_{11} + C_{12} &=& 4(f_{l} - f_{t^{\prime}} - f_{t} )/a \\
C_{11} - C_{12} &=& (f_{l} + 5f_{t^{\prime}} + f_{t} )/a \\
 C_{44} &=& (f_{l} + f_{t^{\prime}} + 2f_{t} )/a \end{eqnarray*}

The values of C$_{11}$, C$_{12}$ and C$_{44}$ are taken from Leibfried and Breuer \cite{Leibfried} (given in Table I).\\
It has been observed that the spectral functions and the structure factors for Ni$_{88}$Cr$_{12}$ has strong evidence of branch dependent widths as also
asymmetry in certain directions. This lends credence to our belief that force-constant disorder leads
to both asymmetry and strong wave-vector and frequency dependence of the line-shapes.

\begin{figure}[b]
\includegraphics[width=8.5cm,height=7cm]{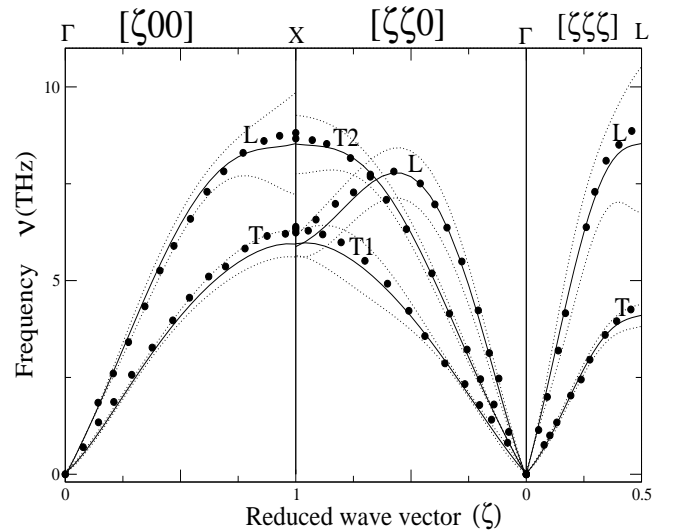}
\caption
{Dispersion Curves (\ frequency $\nu$ vs. reduced wave vector $\zeta$\ ) for Ni$_{88}$Cr$_{12}$ calculated from recursion (solid line). The force constants used are given in the text. The filled circles are the 2 CPA results \cite{thesis} . In all the three panels the thin dotted lines span the FWHM's . }
\label{dispNiCr}
\end{figure}

\begin{figure}
\includegraphics[width=8cm,height=8cm]{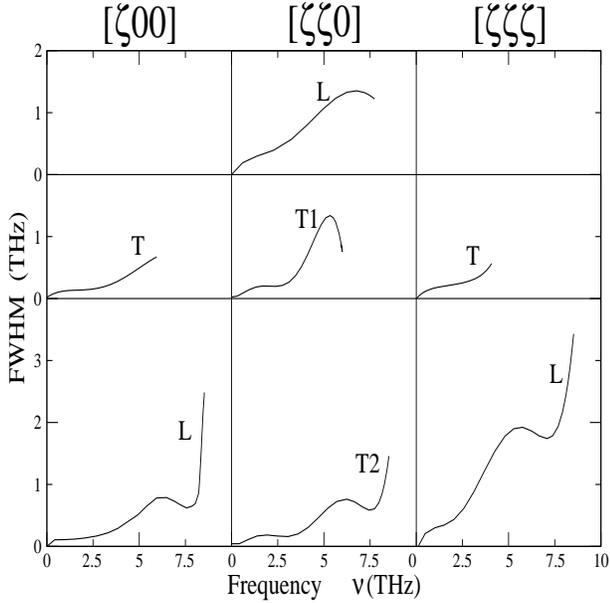}
\caption
{Full widths at half-maximum for the NiCr alloy as function of frequency for different directions
in k-space and different modes.}
\label{widNiCr}
\end{figure}

The influence of force constant disorder may be demonstrated more prominently in the dispersion curves and widths.
 In figure {\ref{dispNiCr}}, we display the dispersion curves along with the FWHM's using the force constants of {Table~\ref{table1}}. The procedure has already been discussed in the previous section. 
These dispersion curves compare well with the experimental results \cite{Bosi} as well as the
2 CPA results of Mookerjee and Singh (\cite{thesis}) (filled circles). The dotted lines span the calculate FWHM's. It should be noted that in the low frequency region, the widths are small but start to become significant as the phonon frequency increases. The widths are comparatively larger in the [$\zeta 0 0 $] L, [$\zeta\zeta\zeta$] L and [ $\zeta\zeta 0$] T2 branches for high $\zeta$-values. Looking at the dispersion curves, one should notice that the behaviour of the natural widths were somehow complemented in the behaviour of the frequencies. There is little evidence of resonances. This is expected, since clear cut
resonances are characteristics of large mass disorders only. 

In figure \ref{widNiCr} we show the FWHM as a function of frequency. It is clear that there
is strong evidence of mode and {$\bf$ {k}}-dependence. The FWHMs are very large and asymmetric for the longitudinal
modes near the band edge frequencies.

\begin{figure}
\includegraphics[width=8.5cm,height=7.0cm]{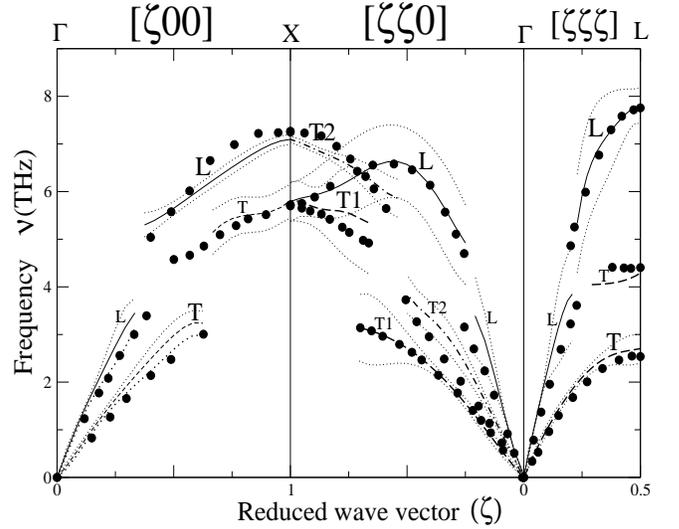}
\caption
{Dispersion Curves (\ frequency $\nu$ vs. reduced wave vector $\zeta$\ ) for Ni$_{50}$Pt$_{50}$ calculated from recursion . The force constants used are given in the text. The solid lines are the L-branch in all the three panels, the dashed lines are the T-branch in the left and right panels. In the [$\zeta$ $\zeta$ 0] direction the dashed line indicate the T1 branch while the dot-dashed line indicate the T2 branch. The filled circles are the ICPA results \cite{subhra}. In all the three panels the thin dotted lines span the FWHM's.  }
\label{dispNiPt}
\end{figure}

It is obvious from the above discussions that the force constant disorder plays a significant role in Ni$_{88}$Cr$_{12}$ ; and a theory capturing only mass disorder effect (\ e.g. like CPA\ ) fails to provide various essential features.  
 
\subsection{The Ni$_{50}$Pt$_{50}$ alloy}

Being encouraged by the right trend of theoretical results toward the experimental results in the Ni$_{55}$Pd$_{45}$ and Ni$_{88}$Cr$_{12}$ alloys, where either of the two disorders -- diagonal and off-diagonal dominates, we now apply our formulation to NiPt alloys where both disorders are predominant. The mass ratio m$_{Pt}$/m$_{Ni}$ is  3.3 (quite large compared to previous alloys) and the nearest neighbour
 force constants of Pt are on an average 55\% larger than those in Ni. Tsunoda {\em et.al} \cite{ts}  have studied this system thoroughly covering a wide range of concentration ($x$=0.05,\ $x$=0.3,\ $x$=0.5) 
by inelastic neutron scattering. In our case, we have considered x=0.5 
where we expect the disorder induced scattering to have the strongest effect.

 In this case, the spectral functions as well as the structure factors show few extra features : Even in [$\zeta$ 0 0 ]L, [$\zeta$ 0 0 ]T and 
[$\zeta$ $\zeta$ 0]T2 modes,  unlike the previous two cases both the functions 
have one usual well defined peak ( observed more clearly in the middle-regime of 
the Brillouin zone ) along with a weakly defined peak with no gap in between. 
The occurrence of such a weakly defined peak is due to the inclusion of force constant disorder.
Ghosh {\em et.al} (\cite{subhra}) have argued that it is entirely because of the off-diagonal
disorder in the force constants. We refer the reader to their paper for the detailed
arguments. Here we note that the feature is equally well reproduced in our augmented space
recursive technique as well. why this should be so ?
The effect of force constant disorder can be understood more clearly by looking at the dispersion curves and widths.

\begin{figure}[t]
\includegraphics[width=8cm,height=8cm]{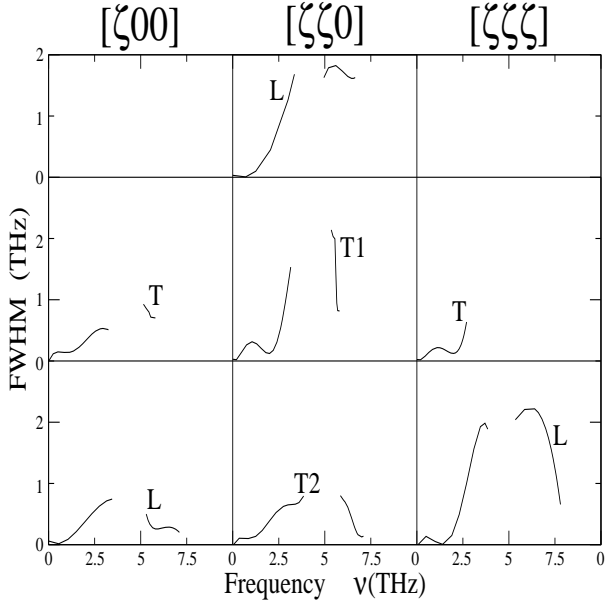}
\caption
{Full widths at half-maximum for the NiPt alloy as function of frequency for different directions
in k-space and different modes including the scattering length fluctuation.}
\label{NiPtwidth}
\end{figure}
 In figures \ref{dispNiPt} and \ref{NiPtwidth} we display the dispersion curves and widths respectively obtained in the recursion using the force constants as given in the text. The solid lines are the L-branch in all the three panels, the dashed lines are the T-branch in the left and right panels, In the middle panel the dashed line indicate the T1-branch while the dot-dashed line indicate theT2-branch. We have used the same parametrization of the force constants as used by Ghosh {\em{et al}}. These dispersion curves (solid lines) are compared with that calculated in the ICPA (filled circles) by Ghosh {\em et.al} \cite{subhra}. 
 The aim of this work was to establish
the ASR as a computationally fast and accurate method for phonon calculations for random alloys. Our
ultimate  goal is to obtain the force constants from first principles calculations.  

For all the three panels the thin dotted lines indicate the FWHM's. Unlike the previous two cases,
 the dispersion curves in this case have very different characteristic features. 
The splitting of the curves in all the three symmetry directions is the main feature. This is due to
strong resonances, a feature of large mass disorder.  Also as one can see that near the resonances
 (around 4 THz) the FWHM becomes very large. Tsunoda finds this resonance near 3.8 THz, while the
CPA gives a rather lower value of 3 THz. In addition to these features , It has been observed that around 7 THz, the structure factor has a third small peak split from the main branch. This evidence of a weak resonance was also speculated by Mookerjee and Singh (\cite{ms1}).
The overall agreement of our dispersion curves with those calculated in the ICPA is good .

\section{Discussion and conclusions}

We have set up the augmented space recursion in reciprocal space for the study of phonon dispersion and disorder induced line-widths and line-shapes for random binary alloys. The technique takes into account
both diagonal disorder in the masses,  the off-diagonal disorder in the force constants and the
environmental disorder in the diagonal term of the force constants arising out of the sum rule. The 
approximation involving termination of continued fraction expansions of the Green function retains the
essential Herglotz analytic properties. We have applied the method to three classes of alloys : NiPd
where mass disorder dominates, NiCr where force constant disorder dominates and NiPt where both dominate.
Wherever possible we have compared our results with neutron scattering data as well as the most sophisticated mean field theory recently proposed by Ghosh {\em et.al} (\cite{subhra}). Both qualitatively and quantitatively
our results agree well with the available data. We propose the technique as a computationally efficient
method for the study of phonons in disordered systems. Our approach here made no attempt to obtain
the force constant themselves from first principles, but rather resorted, as others did earlier, to fitting
them from experimental data on the constituent metals. Our future endeavor would be to rectify this, and
attempt to obtain the dynamical matrix itself from more microscopic theories.
\section{\bf Acknowledgements}
A.A. would like to thank the Council of Scientific and Industrial Research, India, for financial assistance.  
\appendix
\section{The augmented space formalism} 

Let $f(n_R)$ be a function of a random variable $n_R$ , whose binary probability density is given by :

\[ Pr(n_R) \ =\  x_A\ \delta(n_R)\ +\  x_B\ \delta(n_R-1) \]

We may then write :

\[ Pr(n_R) \ =\  -\frac{1}{\pi}\ \Im m \langle\uparrow_R \vert (n_R I - N_R)^{-1}\vert\uparrow_R\rangle \]

Here, the operator $N_R$ acts on a space spanned by the eigenvectors $\vert 0\rangle $ and $\vert 1\rangle$ of
$N_R$, corresponding to eigenvalues 0 and 1 ; $\vert\uparrow_R\rangle =\sqrt{x_A}\vert 0\rangle +\sqrt{x_B}\vert 1\rangle $
is called the {\sl reference state}. Its orthogonal counterpart is 
$\vert\downarrow_R\rangle =\sqrt{x_B}\vert 0\rangle -\sqrt{x_A}\vert 1\rangle $. The representation of $N_R$ in this new basis is :

\[ N_R \ =\  \left( \begin{array}{cc}
                   x_A & \sqrt{x_A x_B} \\
               \sqrt{x_A x_B} & x_B 
                   \end{array} \right) \]

Now,

\begin{eqnarray}
\ll f(n_R)\gg &  = & \int_{-\infty}^{\infty}\  f(n_R)Pr(n_R) dn_R \nonumber\\
 \phantom{\ll f(n_R)\gg}& =& -\frac{1}{\pi}\Im m \int_{-\infty}^{\infty}\ f(n_R)\langle\uparrow_R\vert (n_RI-N_R)^{-1}\nonumber\\
& & \dots \vert\uparrow_R\rangle dn_{R}\nonumber\\
 \phantom{\ll f(n_R)\gg}& =& -\frac{1}{\pi}\Im m \sum_{\lambda=0,1}\sum_{\lambda '=0,1} \int_{-\infty}^{\infty}\ f(n_R)
\langle\uparrow_R\vert\lambda\rangle\nonumber\\
& & \ldots \langle\lambda\vert (n_RI-N_R)^{-1}\vert\lambda '\rangle\langle\lambda'\vert\uparrow_R\rangle dn_{R} \nonumber\\
 \phantom{\ll f(n_R)\gg}& =& \ \sum_{\lambda =0,1}  \langle\uparrow_R\vert\lambda\rangle f(\lambda)\langle \lambda\vert\uparrow_R\rangle\nonumber\\
 \phantom{\ll f(n_R)\gg}& =& \ \langle\uparrow_R\vert \tilde{\mathbf f}\vert \uparrow_R\rangle
\end{eqnarray}

Here $\tilde{\bf f}$ is an operator built out of $f(n_R)$ by simply replacing the variable $n_R$ by the associated operator $N_R$. The
above expression shows that the average is obtained by taking the matrix element of this operator between the {\sl reference state}
$\vert\uparrow_R\rangle$. The full Augmented Space Theorem is a generalization of this for functions of many independent random 
variables $\{n_R\}$.

\section{Operations in augmented reciprocal space}

The main operations of the effective dynamical matrix on a general state $\vert {\bf k}\otimes \{c\}\rangle\ =\ \vert\vert\{c\}\rangle$ in the augmented reciprocal space are given below, where $\{c\} = \{R_{1},R_{2},\ldots , R_{c}\}$ indicates the cardinality sequence with $\vert\downarrow\rangle$ configuration at $R_{1}, R_{2},\ldots , R_{c} $  sites. This is required to implement the recursion procedure :
\begin{widetext}
\begin{eqnarray*}
\left(\rule{0mm}{4mm} \sum_R\ p_R^\downarrow \otimes P_R\right) \parallel \{{\cal C}\}\rangle &\ =\ &
\left(\rule{0mm}{4mm} \sum_R\ p_R^\downarrow \otimes P_R\right)  \frac{1}{\sqrt{N}} \sum_{R'}
\ \exp{(\imath {\mathbf k}\cdot R')} \vert R',\{{\cal C}\}\rangle  \\
 & \ =\ & \parallel \{{\cal C}\}\rangle \ \delta\left(\rule{0mm}{3mm}R_0\ \in\ \{{\cal C}\}\right)
\quad\mbox{($R_0$ is any reference site)} \\
\phantom{x} & & \\
\left(\rule{0mm}{4mm} \sum_R\ {\cal T}_R^{\uparrow\downarrow} \otimes P_R\right) \parallel \{{\cal C}\}\rangle &\ =\ &
\parallel \{{\cal C}\pm R_0\}\rangle \\
\phantom{x} & & \\
\left(\rule{0mm}{4mm} \sum_R\sum_\chi\ \tilde{\cal I}
\otimes T_{R,R+\chi}\right) \parallel \{{\cal C}\}\rangle &\ =\ & 
  \left(\rule{0mm}{4mm} \sum_R\sum_\chi\ \tilde{\cal I}
\otimes T_{R,R+\chi}\right)
  \frac{1}{\sqrt{N}}\sum_{R'}   
\ \exp{(\imath {\mathbf k}\cdot R')} \vert R',\{{\cal C}\}\rangle  \\
 & \ =\ & s({\mathbf k}) \parallel \{{\cal C-\chi}\}\rangle \quad\mbox{($\chi$ is a lattice vector)}
\end{eqnarray*}

\centerline{where   $s({\mathbf k}) = \sum_\chi\ \exp{(-\imath{\mathbf k}\cdot \chi)}$} 

\begin{eqnarray}
 \phantom{x}& &\nonumber\\
\left(\rule{0mm}{4mm} \sum_R\sum_\chi\ \left(p_R^\downarrow +p_{R+\chi}^\downarrow\right)
\otimes T_{R,R+\chi}\right) \parallel \{{\cal C}\}\rangle & = &  
 s({\mathbf k}) \parallel \{{\cal C-\chi}\}\rangle \ \left[
\delta\left(\rule{0mm}{4mm}R_0 \in \{{\cal C-\chi}\}\right) + \delta\left(R_0+\chi \in \{{\cal C-\chi}\}\right)\right] \nonumber\\
 \phantom{x}& & \nonumber\\
\left(\rule{0mm}{4mm} \sum_R\sum_\chi\ \left({\cal T}_R^{\uparrow\downarrow} +{\cal T}_{R+\chi}^{\uparrow\downarrow}
\right)
\otimes T_{R,R+\chi}\right) \parallel \{{\cal C}\}\rangle & = & 
 s({\mathbf k}) \left[\rule{0mm}{4mm}\parallel \{{\cal C-\chi}\}\pm R_0\rangle\ +\ \parallel \{{\cal C-\chi}\}\pm(R_0+\chi)\rangle\right]
 \nonumber\\
 \phantom{x}& & \nonumber\\
\left(\rule{0mm}{4mm} \sum_R\sum_\chi\ \left(p_R^\downarrow p_{R+\chi}^\downarrow\right)
\otimes T_{R,R+\chi}\right) \parallel \{{\cal C}\}\rangle & = & 
 s({\mathbf k}) \parallel \{{\cal C-\chi}\}\rangle \ \left[
\delta\left(\rule{0mm}{4mm}R_0 \in \{{\cal C-\chi}\}\right) \delta\left(R_0+\chi \in \{{\cal C-\chi}\}\right)\right] \nonumber\\
 \phantom{x}& & \nonumber\\
\left(\rule{0mm}{4mm} \sum_R\sum_\chi\ \left({\cal T}_R^{\uparrow\downarrow} {\cal T}_{R+\chi}^{\uparrow\downarrow}
\right)
\otimes T_{R,R+\chi}\right) \parallel \{{\cal C}\}\rangle & = &  
 s({\mathbf k}) \left[\rule{0mm}{4mm}\parallel \{{\cal C-\chi}\}\pm R_0\pm(R_0+\chi)\rangle\right]
 \nonumber\\
 \phantom{x}& & \nonumber\\
\left(\rule{0mm}{4mm} \sum_R\sum_\chi\ \left(p_{R+\chi}^\downarrow {\cal T}_R^{\uparrow\downarrow}+p_R^\downarrow
{\cal T}_{R+\chi}^{\uparrow\downarrow} \right) \ldots
  \otimes T_{R,R+\chi}\right) \parallel \{{\cal C}\}\rangle &  = &  
  s({\mathbf k}) \left[\rule{0mm}{4mm}\parallel \{{\cal C-\chi}\}\pm R_0\rangle\ \delta\left(\rule{0mm}{3mm}(R_0+\chi)\in\{{\cal C-\chi}\}\right) + \ldots\right.\nonumber\\
& &  \ldots\left.\parallel \{{\cal C-\chi}\}\pm(R_0+\chi)\rangle\ \delta\left(\rule{0mm}{3mm}R_0\in\{{\cal C-\chi}\}\right)  
\rule{0mm}{4mm}\right]   
\end{eqnarray}
\end{widetext}
We note that all operations involve only manipulations of the configuration part of the state \cite{k-space} (\ i.e. manipulations of the cardinality sequence only\ ). The operation of the effective dynamical matrix thus entirely takes place in the configuration space and the calculation does not involve the real space $\cal H$ at all. This is an enormous simplification over the standard augmented space recursion described earlier \cite{cardinality}, where the entire reduced real space part as well as the configuration part was involved in the recursion process. Since one can efficiently store the configurations in bits of words so now the calculation becomes much simpler. These operations finally involve simple bit manipulation routines. 

It is interesting to note that the second operation in the above list creates a new configuration. In the next step of recursion
the third operation translates the entire operation by lattice translations $\{\chi\}$. The cluster of configurations thus `travel' across the
lattice as recursion proceeds.

\section{Terminators}

The recursive calculation described earlier gives rise to a set of continued fraction coefficients
$\{a_n,b_n\}$. In any practical calculation we can go only upto a finite number of steps, consistent
with our computational process. In case the coefficients converge, i.e. if $\vert a_n -a\vert\le\epsilon$,
$\vert b_n -b\vert\le\epsilon$ for $n\ge N$, we may replace $\{a_n,b_n\}$ by $\{a,b\}$ for all $n\ge N$.
In that case the asymptotic part of the contiued fraction may be analytically summed to obtain :

\[ T(E) = (1/2)\left( E-a-\sqrt{(E-a)^2-4b^2}\right)\]  which gives a continuous spectrum $a-2b\le E\le a+2b$.
Since the terminator coefficients are related to the band edges and widths, a sensible criterion for the choice of
these asymptotic coefficients is necessary, so as not to give arise to spurious structures in our calculations.
Beer and Pettifor \cite{bp} suggest a sensible criterion : given a finite number of coefficients, we must choose 
$\{a,b\}$ in such a way so as to give, for this set of coefficients, the minimum bandwidth consistent with no loss of
spectral weight from the band. This criterion is easily translated into mathematical terms. The delta functions that would carry weight out of the band must then be situated exactly at the band edges. We thus demand that the continued fraction diverge simulatneously at both the top and the bottom of the band.

At the band edges :  $T(a\pm2b)=\pm b$ so,
\begin{widetext} 
\[
\ll G(a\pm 2b)\gg = 
 \frac{b_1^2/4}{\displaystyle \pm b - \frac{1}{2}(a_{1}-a) - \frac{b_{2}^{2}/4}
{\displaystyle \pm b -\frac{1}{2}(a_2-a)-{\frac{b_3^2/4}{\displaystyle\ddots\phantom{xx}_{\displaystyle \frac{b_N^2/2}{\pm b-(a_N-a)}}}}}}
\]
\end{widetext}

For a given $a$, the (N+1) eigenvalues of the finite tridiagonal matrix :

\[ \left( \begin{array}{ccccc}
\frac{1}{2}(a_1-a) & \frac{1}{2}b_2 & 0 & \ldots & 0 \\
\frac{1}{2}b_2  & \frac{1}{2}(a_2-a) & \frac{1}{2}b_3 & \ldots & 0 \\
0 & \frac{1}{2}b_3 & \ldots & \ldots & 0 \\
\ldots & \ldots& \ldots & \ldots & \frac{1}{\sqrt{2}}b_N \\
\ldots & \ldots& \ldots & \frac{1}{\sqrt{2}}b_N & (a_N-a) \\
\end{array}\right)\]

are values at which the Green function diverges. The maximum and minimum of this set of
eigenvalues are those values of b for which spectral weight has just split off from the band.
Thus  our choice of $a$ is that value for which the maximum eigenvalue is the largest and
the minimum the smallest.  Since the continued fraction involves $b^2$ then,
\[ b_c = \sup_{\{ a\}}\ b_{max}(a_c) = \inf_{\{a\}}\ \vert b_{min}(a_c)\vert\]

With this choice the terminator $T(E)$ has all the Herglotz properties required. Luchini and Nex
\cite{ln} further modified this by replacing the ``butt joining" $\{a_n,b_n\}$ to $a,b$ by a
smooth linear interpolation :

\[  \hat{a}_n, \hat{b}_n = \left\{ \begin{array} {ll}
	   a_n,b_n & n<n_1 \\
(a_n(N-n)+a(n-n_1)]/(N-n_1) & n_1\leq n\leq N \\
(b_n(N-n)+b(n-n_1)]/(N-n_1) & n_1\leq n\leq N \\
a,b & n>N
\end{array} \right. \]

They argued that most of the possible spurious structures are removed by such interpolation.
In our work we have used these two ideas to estimate the terminator.\\


\end{document}